%% file: JungKim_JCN.tex
\documentclass[10pt, twocolumn]{IEEEtran}
\usepackage{fancyhdr}
\usepackage{epsfig}
\usepackage{threeparttable}
\usepackage{epsf,epsfig}
\usepackage{amsthm}
\usepackage{amsmath}
\usepackage{amssymb}
\usepackage{amsfonts}
\usepackage[noadjust]{cite}
\usepackage{dsfont}
\usepackage{subfigure}
\usepackage{color} 
\usepackage{enumerate}
\usepackage{commath}

\include{header}

\begin{document}

\title{Bidding, Pricing, and User Subscription Dynamics in Asymmetric-valued Korean LTE Spectrum Auction: A Hierarchical Dynamic Game Approach }

\author{\IEEEauthorblockN{Sang Yeob Jung and Seong-Lyun Kim}
\thanks{Sang Yeob Jung and Seong-Lyun Kim are with the School of Electrical and Electronic Engineering, 
Yonsei University, Seoul 120-749, Korea. (e-mail: \{syjung, slkim\} @ramo.yonsei.ac.kr).
Part of this paper has been presented in IEEE WiOpt 2014. }}

\maketitle

\begin{abstract}

The tremendous increase in mobile data traffic coupled with fierce competition in wireless industry brings about spectrum scarcity and bandwidth fragmentation. 
This inevitably results in asymmetric-valued LTE spectrum allocation that stems from different timing for twice improvement in capacity between competing operators, given spectrum allocations today. 
This motivates us to study the economic effects of asymmetric-valued LTE spectrum allocation.
In this paper, we formulate the interactions between operators and users as a hierarchical dynamic game framework, where
two spiteful operators simultaneously make spectrum acquisition decisions in the upper-level first-price sealed-bid auction game, and dynamic pricing decisions in the lower-level differential game, taking into account user subscription dynamics. 
Using backward induction, we derive the equilibrium of the entire game under mild conditions.
Through analytical and numerical results, we verify our studies by comparing the latest result of LTE spectrum auction in South Korea, which serves as the benchmark of asymmetric-valued LTE spectrum auction designs. 
\end{abstract}


\section{Introduction}
\subsection{Motivation}
With the proliferation of smartphones, tablets, and ever more data hungry applications,
the demand for mobile data traffic continues to grow dramatically. 
According to a Cisco report, global mobile data traffic will increase
10-fold between 2014 and 2019 \cite{Cisco:2014}. 
As a remedy, the numerous technology enhancements of LTE (i.e., LTE-Advanced) have been proposed, yet 
this technology evolution alone is not sufficient \cite{America:2013, Huang:2013}. 
The need for additional spectrum is critical to address the explosive data challenges successfully.
However, wireless spectrum is a scarce resource, and thus has been tightly regulated by allocating spectrum blocks to licensed cellular operators \cite{Marcus:2015}. Moreover, given spectrum allocations today, it is less likely to exploit a contiguous bandwidth wider than 20 MHz or even a 20MHz contiguous bandwidth \cite{Mikio:2010}. 
These spectrum scarcity and bandwidth fragmentation pose increasing challenges of additional spectrum allocation, from a regulator's perspective.

Our study is motivated by the latest LTE spectrum auction in South Korea on the basis of the above considerations \cite{Moody:2013}. 
The key issue was whether or not Korea Telecom (KT) secures 10 MHz of contiguous spectrum in the 1.8 GHz band.
Because of KT's current 10 MHz downlink carrier in this band, 
KT could directly launch the wideband LTE services (i.e., up to 150 Mbps in the downlink) at little expense, 
offering faster connections compared with the normal LTE (i.e., up to 75 Mbps in the downlink).
This is because LTE Release 8/9 can support a maximum of 20 MHz of contiguous spectrum. 
On the other hand, 
the other operators requested the KCC to exclude KT from bidding on the contiguous spectrum block to ensure fair competition.
In order to provide the double-speed LTE services like KT, 
they should deploy the inter-band non-contiguous carrier aggregation (CA) (i.e., LTE-Advanced services), 
which requires huge investments as well as some deployment time. 
In light of these conflicting views, 
the allocation of the new spectrum resources is not purely a technical issue, 
but economic and regulatory considerations should be taken into account as well.

Spectrum auctions have been widely used in wireless communication. 
Most of prior works have assumed that bidders (i.e., operators) are \textit{self-interested}, i.e., they only maximize 
their own profits regardless of the profits of other bidders \cite{Milgrom:1982}--\cite{Yu:2014}.
However, considerable mismatches between the theory of self-interest and the outcome of the real-world auction have been observed \cite{Illing:2003}. This can be explained by a \textit{spite motive}, which is the preference to deteriorate the profits of their competitors \cite{Morgan:2003}. 
In a highly competitive wireless industry, for instance, 
the loss for one side may entail the corresponding gain for the other side. 
Therefore, operators will intend to maximize their own profit, as well as minimize the profits of their competitors in the auction for improving their own standing, which inspired our work to model and analyze the bidding behavior of such spiteful operators.

\subsection{Contributions of this paper}
This paper studies bidding and dynamic pricing competition between two spiteful/competing operators, taking into account user subscription
dynamics. 
Given that asymmetric-valued LTE spectrum blocks are auctioned off to them, 
we formulate the interactions between two operators and users as a hierarchical dynamic game framework. 
In the upper level, two spiteful operators compete in a first-price sealed-bid auction with considering their current spectrum holdings. 
Different from the standard auction game in the existing literature, 
each operator maximizes the weighted difference of his own profit to that of his competitor.  
In the lower level, two competing operators optimally set their dynamic pricing strategies to maximize their long-term revenues,
considering user subscription dynamics with the newly allocated spectrum. 
Unlike the traditional static game model \cite{Jung:2014}, we formulate a (noncooperative) differential game 
and derive a closed-loop Nash equilibrium to capture the price dynamics and the corresponding user subscription dynamics.
Using backward induction, we derive the equilibrium of the entire game under mild conditions.

The contributions of this paper are summarized follows: 
\begin{itemize}
\item We propose a hierarchical dynamic game framework to study bidding, dynamic pricing, 
and user subscription dynamics. 
This framework can be used as a research tool to analyze the impact of spectrum allocation on market structure.
\item We highlight the impact of different timing to launch the double-speed LTE services between two operators on price dynamics and user subscription dynamics. We further investigate the effects of users' net switching costs, playing a role in a \textit{subsidy} for the market share leader, and a \textit{tax} for the market share followers. Based on this, we examine the asymmetric values of contiguous spectrum and non-contiguous spectrum blocks.
\item We study the bidding behavior of spiteful operators and the resultant profits of them. We verify our studies by comparing 
the latest results of LTE spectrum auction in South Korea. This provides the design guidelines of asymmetric-valued LTE spectrum auction.
\end{itemize}

The remainder of this paper is organized as follows. 
\textrm{S}ection II presents the system model and the proposed hierarchical dynamic game framework.
\textrm{S}ections III and IV investigate the closed-loop Nash equilibriums taking into account user subscription dynamics in asymmetric and symmetric phases, respectively.  
\textrm{S}ection V analyzes the spiteful bidding strategies and the resultant profits.
\textrm{S}ection VI provides numerical results to study the economic effects of asymmetric-valued LTE spectrum allocation, followed by
concluding remarks in Section VII.

\begin{figure}[t]
\centering
\includegraphics[angle=0, width=3.6in]{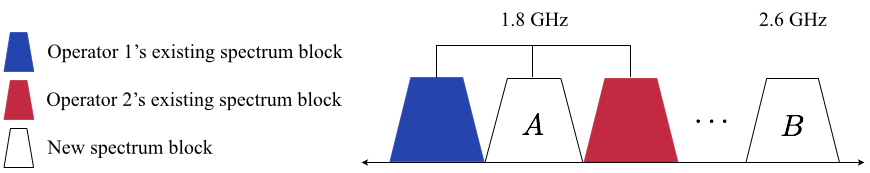}
\caption{System model for spectrum auction.}
\vskip 5pt
\end{figure}

\section{System Model and Game Formulation}
\subsection{System Model}
We consider two operators, say 1 and 2, bidding for two spectrum blocks $A$ and $B$ in a first-price sealed-bid auction
as shown in Fig. 1. In this auction, the two operators simultaneously submit their bids in closed envelopes and the operator with the highest 
bid wins, and pays its bid for the spectrum block. Note that $A$ and $B$ are the same amount of downlink 10 MHz bandwidth. 
Without loss of generality, we consider only the downlink throughout the paper. 
Note that both operators deploy Frequency Division Duplex LTE (FDD LTE) and provide services to users in the same geography area.

Due to the operators' existing spectrum holdings, 
the timing to launch the double-speed LTE services depends on the results of the LTE spectrum auction.
If $A$ is assigned to operator 1, the double-speed LTE services\footnote{In fact, the peak downlink data rate for a user equipment (UE) category 3 (i.e., the existing LTE devices) is 100 Mbps, and 150 Mbps for a UE category 4 (i.e., LTE-Advanced devices). However,
we assume that the double-speed LTE services are supported to the users for analytical tractability, which does not change the main insights obtained in this paper.} are directly provided to users.  
On the other hand, the other operator 2 who acquires $B$ should deploy the inter-band non-contiguous CA to exploit fragmented 
spectrum, enabling to offer double-speed LTE service to users.
However, it requires some deployment time $T$. 
For notational convenience, 
we assume that operator 1 acquires $A$, which will be relaxed in \textrm{S}ection V.

Since the deployment time $T$ completely alters the nature of the operators' dynamic pricing strategies and the user subscription dynamics, 
we analyze two phases $t < T$ and $ t \geq T$, separately, by defining the following terms. 
\vskip -5pt
\begin{definition} (\text{Asymmetric phase})  
Assume that operator 2 launches double-speed LTE service at time $T$. When $0\leq t < T$, we call
this period asymmetric phase due to the different services provided operators 1 and 2. 
\end{definition}
\begin{definition}
(\text{Symmetric phase}) When $t \geq T$, we call this period symmetric phase because of the same services offered by operators 1 and 2. 
\end{definition}
\vskip -5pt

\begin{figure}[t]
\centering
\includegraphics[angle=0, width=3.3in]{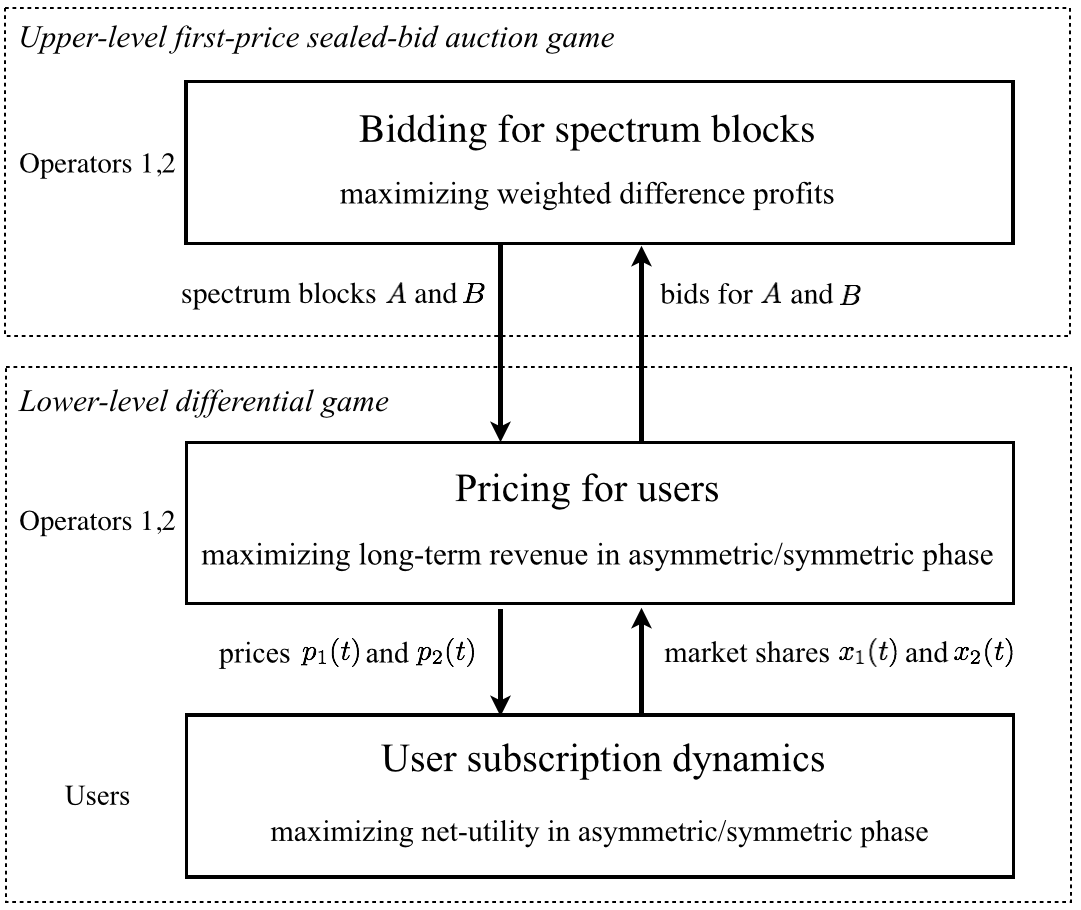}  
\caption{Hierarchical dynamic game framework for bidding, pricing, and user subscription dynamics.} 
\vskip 5pt
\end{figure}

\subsection{A Hierarchical Dynamic Game Framework}
We formulate the interactions between the two operators and the users as a hierarchical dynamic game as shown in \textrm{F}ig. 2. 
The proposed game consists of two levels: an upper-level first-price sealed-bid auction game for bidding competition between 
two spiteful operators, and a lower-level differential game for dynamic pricing competition between them considering 
user subscription dynamics.

\subsubsection{Lower-level differential game} Users dynamically decide whether to stay in their current operator or to switch to the other operator for utility maximization. 
This user subscription dynamics depends on the perceived utilities, the operators' current prices, switching costs, and price subsidies. 
Given the current operators' market shares, the operators optimally set their service prices (i.e., time-dependent) to maximize their long-term revenues with the newly allocated spectrum. 
We model a differential game to completely characterize the operators' dynamic pricing strategies and the corresponding user subscription dynamics
in two phases.

\subsubsection{Upper-level first-price sealed-bid auction game} Two spiteful operators compete in a first-price sealed-bid auction based on 
the estimated revenues in the lower level.
Since the realized profits are tightly coupled across the operators, 
the objective of each operator is maximizing the weighted difference of his own profit to that of his rival,
where the weight $\gamma$ is the spite (or competition) coefficient that denotes the degree of competition.

\section{User Subscription Dynamics and Dynamic Pricing Competition in Asymmetric Phase}

We consider a continuum of users that subscribes to one of the operators based on his or her operator preference. 
This continuum model, a widely used model to analyze the wireless communication industry, reflects 
the reality well if there are a sufficiently large number of users so that a single user is negligible \cite{Chau:2010, Ren:2011}.  
We assume that the total population is normalized to 1. 
Let $x_{i}(t)$ denote the market share of operator $i \in \{1, 2\}$, 
where $x_{i}(0)=x_{i}^{0} \in [0,1]$ is the initial market share of operator $i$.  
We assume full market coverage, i.e., $x_{1}(t) + x_{2}(t) = 1$.
Such assumption approximates the real world in that global mobile-cellular penetration already stands at 96$\%$ \cite{ITU:statistics}.    
We also assume that operators 1 and 2 provide same quality in communication services to the users 
so that they have the same reserve utility $u_{0}$ before spectrum allocation.

\subsection{Lower-level Differential Game in Asymmetric Phase}

\subsubsection{User subscription dynamics model}
In asymmetric phase, the users in operators 1 and 2 obtain different utilities, i.e., 
\begin{equation}
u_{1}(t)=(1+\eta)u_{0}, \quad u_{2}(t)=u_{0},  \quad 0\leq t \leq T,
\end{equation}
where $\eta \in (0,1)$ is a user sensitivity parameter to the double-speed LTE service than existing one. 
It means that the users care more about the data rate as $\eta$ increases. 
The users in operator 2 have more incentive to switch to operator 1 as $\eta$ increases. 
When they decide to change operator 1, however, they face two different types of economic factors: 
switching cost and price subsidy. 
The former that a user has to pay when changing operators is the disutility that a user experiences, 
while the latter that the operators give a subsidy to attract their competitors' users is the utility that a user receives. 
Different users confront with different levels of switching costs and price subsidies.
To model such users' heterogeneity, we assume that the price subsidy $p_{s}$ and the switching cost $c_{s}$ are variables over $[0, \underline{s}]$ 
and $[0, \overline{s}]$, respectively. For simplicity, we further assume that the net switching cost $s=c_{s}-p_{s}$, the difference between the switching cost and the price subsidy, is uniformly distributed in $[-\underline{s}, \overline{s}]$ with $0 <\underline{s} \leq \overline{s}$. When $s \rightarrow -\underline{s}$\footnote{This is reasonable 
because, according to \cite{Shin:2014}, SK Telecom, Korea's largest telecom, 
gave the users who purchased Samsung Galaxy S4s an extra 100 dollars, making it a 'minus phone'.}, the user has more incentive to switch to the other operator while the converse is when $s \rightarrow \overline{s}$.

Now let us focus on how user subscription dynamics works in asymmetric phase. 
In the following analysis, we restrict our attention to the case that the switching from one operator to another in both directions is always possible.
Since the gain of one operator represents a loss to another operator under the full market coverage,
this assumption implies that the two operators are competing fiercely for market share 
by retaining their own users as well as stealing their rival's users. 
A user $k$ in operator $i \in \{1, 2\}$, with net switching cost, $s_{k}$, observes the prices charged by operators 1 and 2 ($p_{1}(t)$ and $p_{2}(t)$). 
A user $k$ in operator $i$ will switch to operator $j \ne i$ if 
\begin{equation}
u_{j}(t) - p_{j}(t) - s_{k} > u_{i}(t) - p_{i}(t). 
\end{equation}
Thus the mass of switching users from operator $i$ to $j$ is 
\begin{equation}
q_{ij}(t)=\mathbb{P}[s_{k}< u_{j}(t)-u_{i}(t) - p_{j}(t) + p_{i}(t)].
\end{equation}
Note that the mass of staying users in operator $i$ $q_{ii}(t)=1- q_{ij}(t)$.

Assuming that $p_{1}(t)-p_{2}(t) \in [\eta u_{0} - \underline{s}, \eta u_{0} + \underline{s}]$ so that $q_{21}(t)$ and $q_{12}(t)$ belong to $(0,1)$,
the net change in operator $i$'s market share in the infinitesimal time interval $(t, t+dt]$, the difference between the mass of switching users from operator $j$ to $i$ 
and the mass of switching users from operator $i$ to $j$, can be expressed as follows:
\begin{equation}
x_{i}(t+dt)-x_{i}(t)=q_{ji}(t)x_{j}(t)dt - q_{ij}x_{i}(t)dt.
\end{equation}
Then each operator's market share rate at time $t$ over $[0, T)$ is  
\setlength\arraycolsep{2pt}\medmuskip=3mu
\begin{eqnarray}
\dot{x}_{1}(t)&=&\frac{\eta u_{0} - p_{1}(t) + p_{2}(t) + \underline{s}}{s_{1}} - \frac{2\underline{s}}{s_{1}}x_{1}(t),\nonumber\\
\dot{x}_{2}(t)&=&-\frac{\eta u_{0} - p_{1}(t) + p_{2}(t) + \underline{s}}{s_{1}} + \frac{2\underline{s}}{s_{1}}(1-x_{2}(t)),
\end{eqnarray}
where $s_{1}=\underline{s}+\overline{s}$.
Note that the rates $\dot{x}_{1}(t)$ and $\dot{x}_{2}(t)$ are functions of the current service prices $p_{1}(t)$ and $p_{2}(t)$, the current market shares $x_{1}(t)$ and $x_{2}(t)$, the user sensitivity parameter $\eta$, and the minimum (or maximum) of users' net switching costs $\underline{s}$ (or $\overline{s}$).

\subsubsection{Operators' pricing model}

Given the user subscription dynamics (5), operators 1 and 2 simultaneously determine their prices $(p_{1}(t), p_{2}(t))$ so that 
their total revenues are maximized over $[0, T)$.

Similarly, each operator's revenue rate at time $t$ in asymmetric phase can be written as follows:
\setlength\arraycolsep{-1pt}
\begin{equation}
r_{i}(t)=p_{i}(t)(x_{i}(t) + \dot{x}_{i}(t)),\; i \in \{1, 2\}.
\end{equation}
Here, the revenue rate is the sum of revenue from current users and revenue gain/loss from new/old users. 

For each operator $i\in \{1,2\}$, the optimal pricing control problem over a finite-time horizon $[0, T)$ can be expressed as follows:
\setlength\arraycolsep{1pt}\medmuskip=1mu
\begin{eqnarray}
\bf{P1}: \;\;\;&&\max_{p_{i}(t)}\; \mathcal{R}_{i}^{AP}(p_{1}(t), p_{2}(t))=\int_{0}^{T}e^{-\rho t}r_{i}(t)dt,\\
&&\;\text{s.t.}\; \dot{x}_{1}(t)=\frac{\eta u_{0} - p_{1}(t) + p_{2}(t) + \underline{s}}{s_{1}} - \frac{2\underline{s}}{s_{1}}x_{1}(t), \\
&&\quad\;\; x_{1}(0)=x_{1}^{0},
\end{eqnarray}
where $\rho$ is the discount rate taking into account the time value of money, and $x_{1}^{0}$ is the initial market share of operator $1$. 
Since $\dot{x}_{2}(t)=-\dot{x}_{1}(t)$ and $x_{2}(0)=1-x_{1}(0)$ given the assumption of full market coverage, 
we can describe the user subscription dynamics constraints by only (8) and (9). 
Note that each operator's pricing strategy depends on not only the competitor's pricing strategy but also, 
the market share state (i.e., the market shares of the two operators) that evolves according to a user subscription dynamic constraint in (8).
This allows us to formulate and analyze the optimization problem $\bf{P1}$ within the differential fame framework \cite{Han:2011}.

\subsubsection{Formulation of Differential Game}

Now we formulate the dynamic pricing competition between the two operators as a finite-horizon differential game as follows.

\begin{itemize}
\item Players: two operators 1 and 2.
\item Strategy space: operator 1 can choose price $p_{1}(t)$ from the continuous and bounded set $\mathcal{P}_{1}=[0, (1+\eta)u_{0}]$
while operator 2 can choose price $p_{2}(t)$ from $\mathcal{P}_{2}=[0, u_{0}]$. This is due to the assumption $x_{1}(t)+x_{2}(t)=1$ so that all users 
are served. 
\item Market share state: the market shares of the two operators constitute the market share state denoted by a vector $\bold{x}(t)=[x_{1}(t), x_{2}(t)]^{T}$, where the state is controlled by the user subscription dynamics constraint in (8). 
\item Payoff function: two operators want to maximize their total discounted revenues over a prespecified time horizon in (7), respectively. 
\end{itemize}

\subsection{Closed-loop Nash Equilibrium in Asymmetric Phase}
For the formulated finite-horizon differential game, we will investigate the two operators' dynamic pricing competition and the corresponding user 
subscription dynamics.
To understand the two operator's dynamic pricing strategies in the differential games literature, 
we first want to point out two main types of strategies: open-loop strategies and closed-loop strategies. 
The open-loop strategies do not involve strategic interaction between the two operators 
through the evolution of the market share state over time and the corresponding adjustment in their prices. 
This means that the two operators announce their pricing strategies at the initial time and commit to them,
regardless of how the market share state $\bold{x}(t)$ evolves.
In this regard, the open-loop strategies are neither time-consistent nor subgame-perfect in general. 
On the other hand, the closed-loop strategies take into account the initial and current market share state, allowing  
the two operators to determine and adjust their pricing strategies as $\bold{x}(t)$ changes.  
Thus, we consider the closed-loop strategies to capture the price dynamics and the corresponding user subscription dynamics.

With the notion of closed-loop strategies, the (subgame-perfect) closed-loop Nash equilibrium is defined as follows:
\vskip 5pt \noindent {\bf Definition 3} \emph{(\text{Closed-loop Nash equilibrium}){\bf.}
A pair of closed-loop strategies $(p_{1}^{\ast}, p_{2}^{\ast}) \in \mathcal{P}_{1}\times \mathcal{P}_{2}$ 
is called a closed-loop Nash equilibrium if the following holds:  
\begin{eqnarray*}\small
\mathcal{R}_{i}^{AP}(p_{i}^{\ast}(\cdot), p_{j}^{\ast}(\cdot))\geq \mathcal{R}_{i}^{AP}(p_{i}(\cdot), p_{j}^{\ast}(\cdot)), 
\end{eqnarray*}\normalsize
for all $p_{i}\in \mathcal{P}_{i}$, and for any initial market share $x_{i}(0)=x_i^{0}\in [0,1]$, $i=1,2, i\ne j$.}
\vskip 1pt \noindent
At a closed-loop Nash equilibrium,  
no operator can increase its revenue by changing its strategy unilaterally, given the current pricing strategy of the other operator.

To obtain the closed-loop Nash equilibrium and the resultant user subscription dynamics, 
we need to first examine the necessary conditions by applying the Pontryagin maximum principle \cite{Dockner:2000}.
To this end, we introduce the Hamiltonian function for operator $i$ defined as:
\begin{equation}
\mathcal{H}_{i}(t)=r_{i}(t) + \lambda_{i}(t)\dot{x}_{1}(t), \quad\ i=\{1, 2\},
\end{equation}
where $\lambda_{i}(t)$ is the co-state variable of operator $i$ associated with 
the market share $x_{1}(t)$. For simplicity, we will drop the time dependence expression $t$ from all variables, unless specified otherwise.

Since the Hamiltonian function for operator $i$ is strictly concave in the price $p_{i}$, 
the necessary conditions for the closed-loop Nash equilibrium provide sufficient conditions for optimality, i.e., 
\medmuskip=1mu \thickmuskip=1mu
\begin{eqnarray}
\dot{x}_{1}&=&\frac{\eta u_{0} - p_{1} + p_{2} + \underline{s}}{s_{1}} - \frac{2\underline{s}}{s_{1}}x_{1}, \;\;\;\;x_{1}(0)=x_{1}^{0},\\
\frac{\partial \mathcal{H}_{i}}{\partial p_{i}}&=&0,\\
\frac{\partial \lambda_{i}}{\partial t}&=&\rho \lambda_{i}-\frac{\partial \mathcal{H}_{i}}{\partial x_{1}} - \frac{\partial \mathcal{H}_{i}}{\partial p_{j}}\frac{\partial p_{j}}{\partial x_{1}},\\
\lambda_{i}(T)&=&0.
\end{eqnarray}
Note that (12) is the maximum condition with respect to the strategy of operator $i$, 
(13) is the adjoint equation to describe the dynamics of the co-state variable, and (14) is the boundary condition for the terminal time of the co-state variable. 
We further note that the term $\frac{\partial \mathcal{H}_{i}}{\partial p_{j}}\frac{\partial p_{j}}{\partial x_{1}}$ in (13) 
affects how the two operators adjust their prices as $\bold{x}(t)$ evolves over time. By solving the above conditions, 
we can obtain the following proposition.

\vskip 2pt \noindent {\bf Proposition 1.} 
{\it Let 
\begin{equation}\small
p_{1}^{\ast}(t,x_{1}^{\ast}(t))=\frac{\eta u_{0} + \overline{s} +2\underline{s} - e_{1}(t) - z(t)}{3} + \frac{s_{2}-k(t)}{3}x_{1}^{\ast}(t),
\end{equation}\normalsize
\small\begin{equation}
p_{2}^{\ast}(t, x_{1}^{\ast}(t))=\frac{-\eta u_{0} + 2\overline{s} +\underline{s} + e_{2}(t)- z(t)}{3} - \frac{s_{2}-k(t)}{3}x_{1}^{\ast}(t),
\end{equation}\normalsize
\small\begin{equation}
x_{1}^{\ast}(t)=e^{\int_{0}^{t}\delta(\xi)d\xi}\left(x_{1}^{0}+ \int_{0}^{t}e^{-\int_{0}^{\tau}\delta(\xi)d\xi}\zeta(\tau)d\tau\right), x_{2}^{\ast}(t)=1-x_{1}^{\ast}(t),
\end{equation}\normalsize
where 
\small\begin{equation}
\lambda_{1}(t)=k(t)x_{1}^{\ast}(t)+e_{1}(t), \quad \lambda_{2}(t)=k(t)x_{2}^{\ast}(t)+e_{2}(t)\nonumber
\end{equation}\normalsize
\small\begin{equation}
k(t)=\frac{\alpha_{1}\left(1- e^{\frac{2(\alpha_{1}-\alpha_{2})}{3s_{1}}(T-t)}\right)}{1- \frac{\alpha_{1}}{\alpha_{2}}e^{\frac{2(\alpha_{1}-\alpha_{2})}{3s_{1}}(T-t)}}, \mu(t)=\frac{13\underline{s}+5\overline{s}+6k(t)}{9s_{1}}+\rho, \nonumber
\end{equation}\normalsize
\small\begin{equation}
z(t)=\frac{2s_{2}}{3(1+\rho)}\left(1- e^{(1+\rho)(t-T)}\right), \zeta(t)=\frac{\eta u_{0} +2a + b +e_{1}(t)+ e_{2}(t)}{3s_{1}},\nonumber
\end{equation}\normalsize
\small\begin{equation}
e_{1}(t)=-e^{\int_{0}^{t}\mu(\xi)d\xi}\left(\int_{t}^{T}e^{-\int_{0}^{\tau}\mu(\xi)d\xi}\nu(\tau)d\tau\right), e_{2}(t)=e_{1}(t)-z(t), \nonumber
\end{equation}\normalsize
\small\begin{equation}
\delta(t)=\frac{2(k(t)-b-2a)}{3s_{1}}, \nu(t)=\frac{\left(2s_{2}+3k(t)\right)\left(z(t)- 2\underline{s}- \overline{s} -\eta u_{0} \right)}{9s_{1}},\nonumber
\end{equation}\normalsize
and $\alpha_{1}$ and $\alpha_{2}$ are the solutions of the quadratic equation 
$\frac{2}{3s_{1}}k(t)^{2} - \left(\frac{11\overline{s}+25\underline{s}+9s_{1}\rho)}{9s_{1}}\right)k(t) + \frac{2}{9}s_{2}=0$. Then, 
($p_{1}^{\ast}, p_{2}^{\ast})$ constitute a closed-loop Nash equilibrium 
and $(x_{1}^{\ast}, x_{2}^{\ast})$ are the corresponding user subscription dynamics of the problem $\bf{P1}$}.
\vskip 10pt
\noindent{\bf Proof.} See Appendix A. \hfill $\blacksquare$ \vskip 5pt

\section{User Subscription Dynamics and Dynamic Pricing Competition in Symmetric Phase}

In the previous section, 
we have studied the dynamic pricing competition of the two operators and the corresponding user subscription dynamics 
in asymmetric phase. In this section, we consider them in symmetric phase
by formulating an infinite-horizon differential game. 
The analysis is similar to the previous section and thus this section is brief.

\subsection{Lower-level Differential Game in Symmetric Phase}
\subsubsection{User subscription dynamics model}
Since operator $2$ launches double-speed LTE service in symmetric phase, 
we assume that the users in operators $1$ and $2$ obtain same utility, i.e., 
\begin{equation}
u_{1}(t)=u_{2}(t)=(1+\eta)u_{0}, \quad t>T.
\end{equation}

Similar to the analysis in Sec. III-A, assuming that $p_{1}(t)-p_{2}(t) \in [-a, a]$ so that $q_{21}(t)$ and $q_{12}(t)$ belong to $(0,1)$, 
each operator's market share rate at time $t$ over $[T, \infty)$ is
\begin{equation}
\dot{x}_{1}(t)=\frac{p_{2}(t)-p_{1}(t)+\underline{s}}{s_{1}} - \frac{2\underline{s}}{s_{1}}x_{1}(t)=-\dot{x}_{2}(t).
\end{equation}

\subsubsection{Operators' pricing model}
Given the user subscription dynamics (19), each operator's revenue rate at time $t$ in symmetric phase can be written as follows:
\begin{equation}
r_{i}(t)=p_{i}(t)(x_{i}(t) + \dot{x}_{i}(t)),\; i \in \{1, 2\}.
\end{equation}
For each operator $i\in \{1,2\}$, the optimal pricing control problem over an infinite-time horizon $[T, \infty)$ can be expressed as follows.
\setlength\arraycolsep{1pt}\medmuskip=1mu
\begin{eqnarray}
\bf{P2}: \;\;\;&&\max_{p_{i}(t)}\; \mathcal{R}_{i}^{SP}(p_{1}(t), p_{2}(t))=\int_{T}^{\infty}e^{-\rho t}r_{i}(t)dt,\\
&&\;\text{s.t.}\; \dot{x}_{1}(t)=\frac{p_{2}(t)-p_{1}(t)+\underline{s}}{s_{1}} - \frac{2\underline{s}}{s_{1}}x_{1}(t),\\
&&\quad\;\;\;\;\; x_{1}(T)=x_{1}^{T},
\end{eqnarray}
where $x_{1}^{T}$ is the initial market share of operator $1$ from the end of asymmetric phase.

\subsubsection{Formulation of differential game}
The dynamic pricing competition between the two operators can be formulated as an infinite-horizon differential game as follows.

\begin{itemize}
\item Players: two operators 1 and 2.
\item Strategy space: operator $i$ can choose price $p_{i}(t)$ from the continuous and bounded set $\mathcal{P}_{i}=[0, (1+\eta)u_{0}]$ for $i=1, 2$.       
\item Market share state: the market share state $\bold{x}(t)=[x_{1}(t), x_{2}(t)]^{T}$ is controlled by the user subscription dynamics constraint in (19). 
\item Payoff function: the two operators want to maximize their total discounted revenues over a prespecified time horizon in (21), respectively. 
\end{itemize}

\subsection{Closed-loop Nash Equilibrium in Symmetric Phase}
To proceed with the analytical solution of the formulated infinite-horizon differential game, 
we now focus on the closed-loop Nash equilibrium solutions and the consequent user subscription dynamics of the problem $\bf{P2}$. 
Recall that at a closed-loop Nash equilibrium, no operator has an incentive to deviate from its strategy after considering the other operator's strategy. 
The closed-loop Nash equilibrium and the consequent user subscription dynamics are given in the following proposition.

\vskip 2pt \noindent {\bf Proposition 2.} 
{\it Let 
\begin{equation}\small
p_{1}^{\ast}(t,x_{1}^{\ast}(t))=\frac{s_{1} + \underline{s}+e_{2}-2e_{1}}{3} + \frac{s_{2}-k}{3}x_{1}^{\ast}(t),
\end{equation} \normalsize
\begin{equation}\small
p_{2}^{\ast}(t,x_{1}^{\ast}(t))=\frac{s_{1}+\overline{s}+2e_{2}-e_{1}}{3} - \frac{s_{2}-k}{3}x_{1}^{\ast}(t),
\end{equation} \normalsize
\begin{equation}\small
x_{1}^{\ast}(t)=\frac{1}{2} + \left(x_{1}^{T} - \frac{1}{2}\right)e^{\frac{2(k-\underline{s}-s_{1})}{3s_{1}}(t-T)}, \quad x_{2}^{\ast}(t)=1-x_{1}^{\ast}(t),
\end{equation} \normalsize
where $k$ is the smallest root of the quadratic equation 
\begin{equation}\small
6k^{2}-\left(11\overline{s}+25\underline{s}+9\rho s_{1}\right)k + 2s_{2}^{2}=0,\nonumber
\end{equation} \normalsize
and 
\begin{equation}\small
e_{1}=\frac{s_{2}}{3(1+\rho)}+ \frac{s_{2}^{2}-3k(2\underline{s}+\overline{s})}{6k -13\underline{s}-5\overline{s}-9\rho s_{1}}, e_{2}=e_{1}-\frac{2s_{2}}{3(1+\rho)}.\nonumber
\end{equation} \normalsize
Then, $(p_{1}^{\ast},p_{2}^{\ast})$ constitute a closed-loop Nash equilibrium 
and $(x_{1}^{\ast}, x_{2}^{\ast})$ are the corresponding user subscription dynamics of the problem $\bf{P2}$}.
\vskip 10pt
\noindent{\bf Proof.} See Appendix B. \hfill $\blacksquare$ \vskip 5pt

\section{Bidding Competition}

In the upper level, 
the two spiteful operators compete in a first-price sealed-bid auction 
where asymmetric-valued spectrum blocks $A$ and $B$ are auctioned off to them. 
For fair competition, each operator is constrained to lease only one spectrum block (i.e., $A$ or $B$). 
We assume that the regulators set the reserve prices $c^{A}$ and $c^{B}$ to $A$ and $B$, respectively. 
Note that reserve price is the minimum price to get the spectrum block. 
Since $A$ is the high-valued spectrum block, 
we further assume that the two spiteful operators are only competing on $A$ to provide double-speed LTE service to 
their users earlier.

Based on the projected total revenues of the two operators in the lower level, 
operators 1 and 2 bid $A$ independently as $b_{1}$ and $b_{2}$, respectively. 
In this case, $B$ is assigned to the operator who loses in the auction as the reserve price $c^{B}$. 
Since this operator should make huge investments to double the existing LTE network capacity compared 
to the other MNO, we also assume the only operator who acquires $B$ incurs the investment cost $c^{BS}$.

For notational convenience, we have assumed that operator 1 procures $A$ throughout the paper.
Without loss of generality, however, we can relax this assumption easily after some algebraic manipulations.   
The projected aggregate revenues of the two operators over the entire time-horizon $[0, \infty]$ can be expressed as follows:
\begin{itemize}
\item Assuming that operators 1 and 2 acquire $A$ and $B$ respectively, each operator's projected aggregate revenue during the entire period is 
\begin{eqnarray}
\mathcal{R}_{1}^{A}=\mathcal{R}_{1}^{AP}+\mathcal{R}_{1}^{SP}, \quad \mathcal{R}_{2}^{B}=\mathcal{R}_{2}^{AP}+\mathcal{R}_{2}^{SP},
\end{eqnarray}
where $\mathcal{R}_{i}^{AP}$ and $\mathcal{R}_{i}^{SP}$ are described in (7) and (21), respectively, for $i=1, 2$.
\item Assuming that operators 1 and 2 procure $B$ and $A$ respectively, each operator's projected aggregate revenue during the entire period is 
\begin{eqnarray}
\mathcal{R}_{1}^{B}=\mathcal{R}_{1}^{AP^{'}}+\mathcal{R}_{1}^{SP^{'}},\quad \mathcal{R}_{2}^{A}=\mathcal{R}_{2}^{AP^{'}}+\mathcal{R}_{2}^{SP^{'}},
\end{eqnarray}
where $\mathcal{R}_{i}^{AP^{'}}$ and $\mathcal{R}_{i}^{SP^{'}}$ for $i=1, 2$ can be obtained by reformulating the problems $\bf{P1}$ and $\bf{P2}$, and following the same steps as in Appendices $A$ and $B$. Details are omitted for brevity. 
\end{itemize}

When asymmetric-valued spectrum blocks are allocated to the operators, 
there is a trade-off between self-interest and spite. 
To illustrate this trade-off, we first restrict ourselves to the case where the spite is not present. 
If operator $i$ is $\emph{self-interested}$, his objective function is as follows:
\setlength\arraycolsep{5pt}\medmuskip=1mu\thickmuskip=3mu
\begin{eqnarray}
\Pi_{i}(b_{i}, b_{j})=\left[\mathcal{R}_{i}^{A} - b_{i} \right]\cdot I_{b_{i}\geq b_{j}} + \pi_{i}^{B}\cdot I_{b_{i}\leq b_{j}},  i=1,2, i\ne j,
\end{eqnarray}
where $I$ is the indicator function and $\pi_{i}^{B}=\mathcal{R}_{i}^{B}- c^{B}-c^{BS}$ is the profit from $B$ for $i=1, 2$.
 This is the standard auction framework in that operator $i$ maximizes his own profit without
considering the profit of his competitor.

In the real world, however, there are observations that some operators are $\emph{completely malicious}$ \cite{Illing:2003}. 
If operator $i$ is completely malicious, his objective function can be changed as follows:
\setlength\arraycolsep{4.5pt}\medmuskip=1mu\thickmuskip=3mu 
\begin{eqnarray}
\Pi_{i}(b_{i}, b_{j})=-\pi_{j}^{B}\cdot I_{b_{i}\geq b_{j}} - \left[\mathcal{R}_{j}^{A}- b_{j}\right]\cdot I_{b_{i}\leq b_{j}}, i=1,2,  i\ne j,
\end{eqnarray}
It means that operator $i$ only intends to minimize the profit of operator $j$.

Departing from the standard auction framework, 
our model incorporates this strategic concern in that each spiteful operator cares about maximizing 
the weighted difference of his own profit to that of his competitor.
Combining (29) and (30), we define each operator's objective function as follows.

\vskip 10pt \noindent {\bf Definition 4}{\bf.}\emph{
Assume that two spiteful operators (i.e., $i=1, 2$ and $i\ne j$) compete in a first-price sealed-bid auction.
The objective function that each operator tries to maximize is given by: 
\setlength\arraycolsep{2pt}
\begin{eqnarray}
\Pi_{i}(b_{i}, b_{j})=&\;&\left[(1-\gamma)(\mathcal{R}_{i}^{A} - b_{i}) -\gamma \pi_{j}^{B} \right]\cdot I_{b_{i}\geq b_{j}}\nonumber\\
&+&\left[(1-\gamma)\pi_{i}^{B} -\gamma\left(\mathcal{R}_{j}^{A}- b_{j}\right)\right]\cdot I_{b_{i}\leq b_{j}},
\end{eqnarray}
where $I$ is the indicator function, $\pi_{i}^{B}=\mathcal{R}_{i}^{B}- c^{B}-c^{BS}$ is the profit from $B$, and $\gamma\in [0, 1]$ is a parameter called the spite (or competition) coefficient.}

Note that (31) will be used to determine the bidding strategies of two spiteful operators, which does not affect the realized profits of the two operators. 
As described, operator $i$ is self-interested and only intends to maximize his own profit when $\gamma=0$. 
When $\gamma=1$, operator $i$ is completely malicious and only tries to minimize the profit of his competitor. 
For given $\gamma\in [0,1]$\footnote{If two operators are asymmetrically spiteful, $\gamma$ can be different for two operators.}, we can derive the optimal bidding strategies that maximize the objective function in Definition 4 as follows.
  
\vskip 10pt \noindent {\bf Proposition 3}{\bf.} {\it In a first-price sealed-bid auction, the equilibrium bidding strategies for two operators 1 and 2 are given by
\setlength\arraycolsep{2pt}
\begin{eqnarray}
b^{\ast}_{1}(\gamma)=\frac{(1-\gamma)(\mathcal{R}_{1}^{A}+c^{A}-\pi_{1}^{B}) + \gamma(\mathcal{R}_{2}^{A}-\pi_{2}^{B})}{2-\gamma},\nonumber\\
b^{\ast}_{2}(\gamma)=\frac{(1-\gamma)(\mathcal{R}_{2}^{A}+c^{A}-\pi_{2}^{B}) + \gamma(\mathcal{R}_{1}^{A}-\pi_{1}^{B})}{2-\gamma}.
\end{eqnarray}}
\vskip 10pt
\noindent{\bf Proof.} See Appendix C.
\hfill $\blacksquare$ \vskip 5pt

\section{Numerical Results}

In this section, several numerical results are presented to investigate the impact of spectrum allocation on market structure, 
and provide insights into the role of the regulator. 
For illustration convenience, we assume that operators 1 and 2 acquire the spectrum blocks $A$ and $B$, respectively, unless otherwise specified.

\begin{figure}[ht]
\centering
 \subfigure[Dynamic pricing strategies of operator 1 and operator 2]{
   \includegraphics[angle=0,width=3.1in] {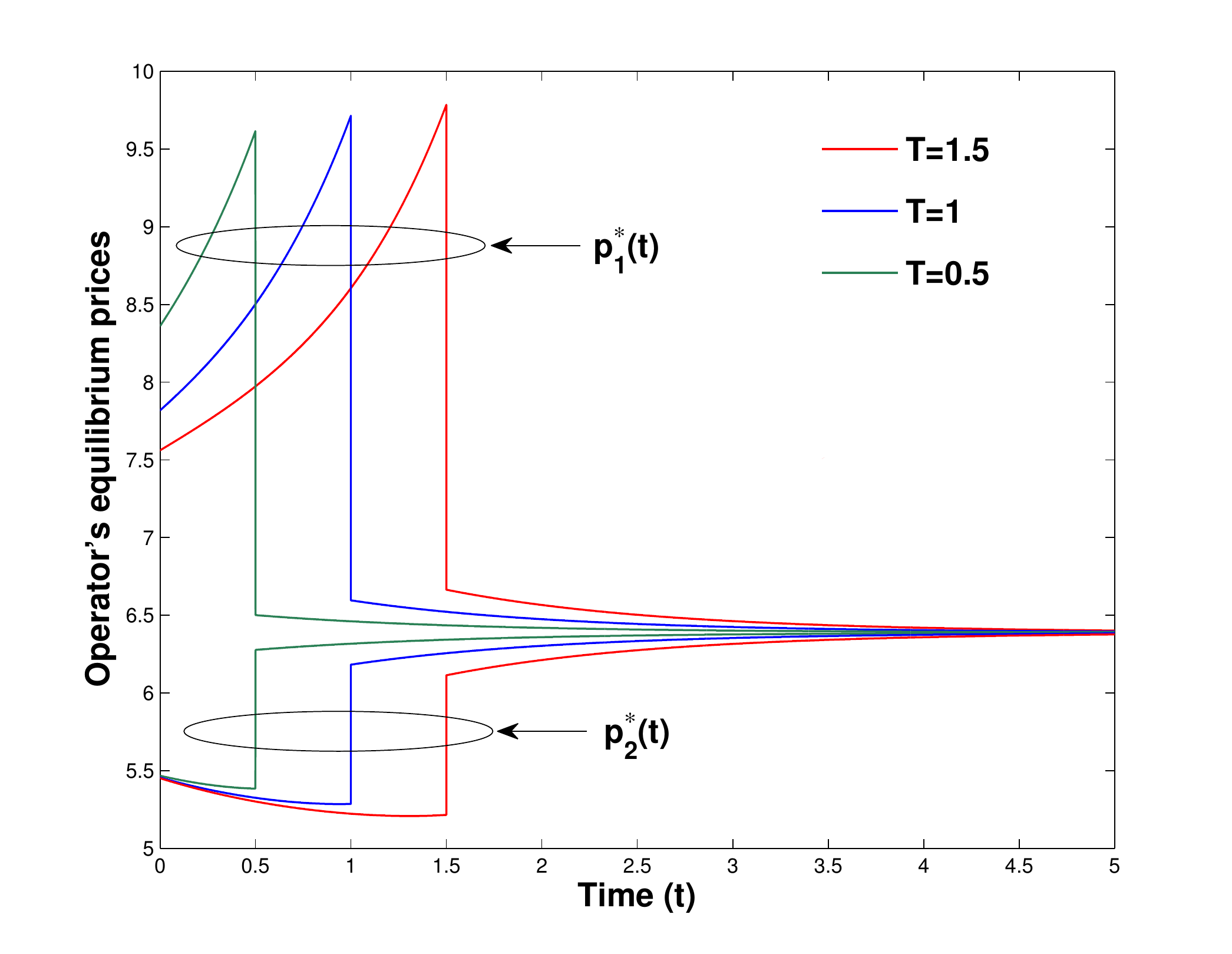} 
 }
\subfigure[User subscription dynamics]{
   \includegraphics[angle=0,width=3.1in] {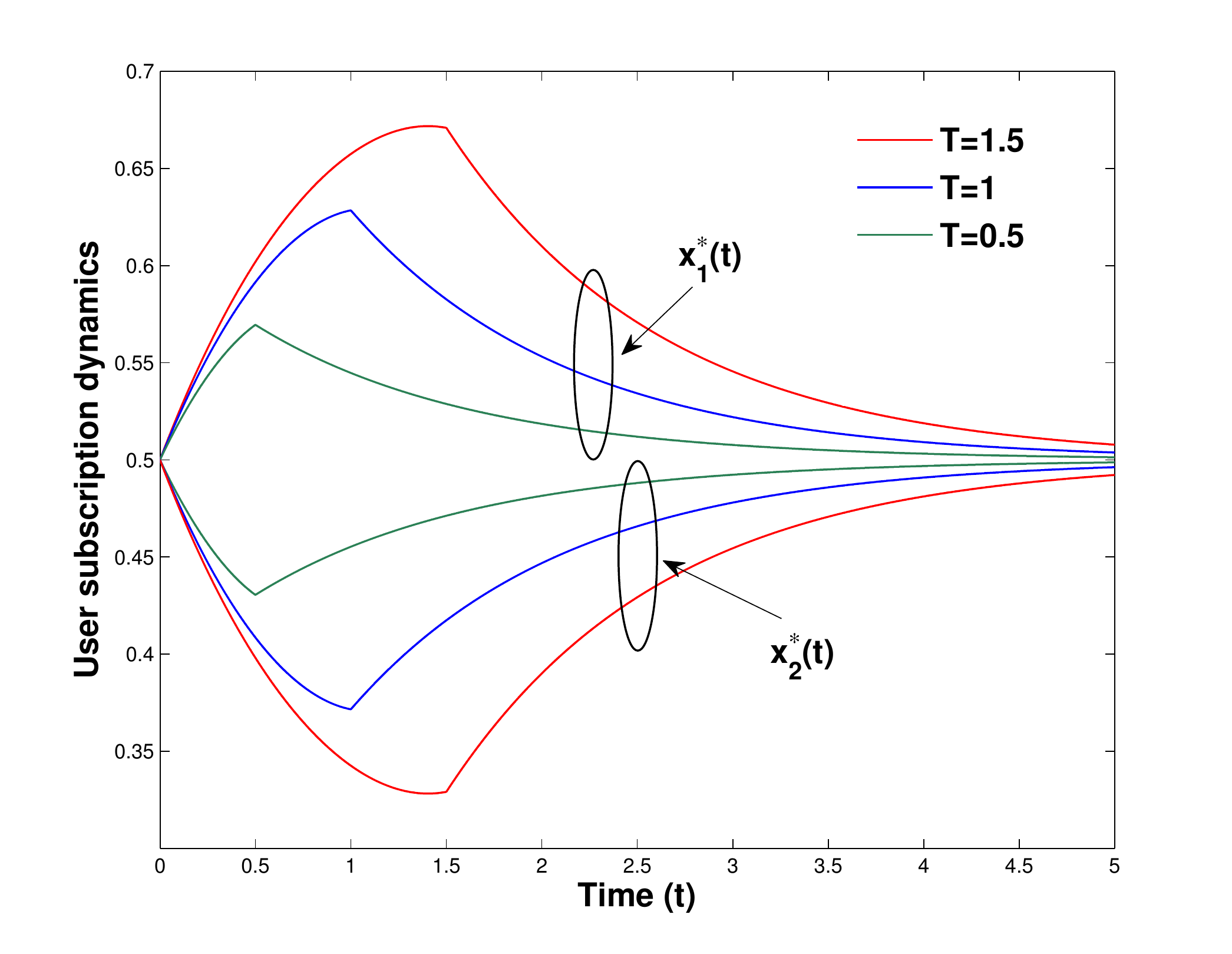}
 }
 \vskip -5pt
\caption{The dynamics of user subscription and the two operators' pricing strategies. Other parameters are $u_{0}=10$, $\eta=0.5$, $\rho=0.5$, $\underline{s}=5$, and $\overline{s}=10$.}
\vskip 1pt
\end{figure}

\subsection {Dynamics of Pricing Competition and User Subscription}
We first study the impact of spectrum allocation on the two operators' dynamic pricing strategies and the the resultant user subscription dynamics.
As a benchmark, we focus on the symmetric market share before spectrum allocation where the initial market state of the two operators
are chosen as $\bold{x}(0)=[0.5, 0.5]$.

\textrm{F}ig. 3 shows the two operators' equilibrium pricing strategies and the corresponding user subscription dynamics under different deployment times. 
In asymmetric phase $(t<T)$, due to the earlier launch of double-speed LTE service, 
we see that operator 1 becomes the market share leader while charging a higher price.  
An interesting observation is that $p_{1}^{\ast}(t)$ and $x_{1}^{\ast}(t)$
increase in $t$, while the reverse is true for operator 2. 
These phenomena are due to the presence of the users' net switching costs. 
As $t$ increases, the more users with the high net switching costs are locked in, allowing operator 1 to optimally raise
$p_{1}^{\ast}(t)$ by the amount of the net switching costs with the increase in $x_{1}^{\ast}(t)$. 
This forces operator 2 to reduce $p_{2}^{\ast}(t)$ but with the decrease in $x_{2}^{\ast}(t)$. 
The users' net switching costs, thus, can be interpreted as a \textit{subsidy} for operator 1 and a \textit{tax} for operator 2. 
It is also worth pointing out that the longer deployment time $T$ is the slower rate of increase of $p_{1}^{\ast}(t)$ for operator 1,
while the faster rate of decrease of $p_{2}^{\ast}(t)$ for operator 2 at the lower initial prices $p_{1}^{\ast}(0)$ and $p_{2}^{\ast}(0)$.
With the longer $T$, operator 1 has more to gain by charging $p_{1}^{\ast}(t)$ less aggressively and attaining $x_{1}^{\ast}(t)$ more aggressively for the same $t$, thereby maximizing the aggregated revenue $\mathcal{R}_{1}^{AP}(T)$ over $[0, T)$. 
On the other hand, operator 2 tries to maximize $\mathcal{R}_{2}^{AP}(T)$ by decreasing $p_{2}^{\ast}(t)$ more aggressively, but losing more $x_{2}^{\ast}(t)$ for the same $t$. 
 
In symmetric phase ($t\geq T$), due to the same services offered by the two operators 1 and 2, 
each operator faces a trade-off between a low price to increase its market share, and a high price to harvest 
its revenue by exploiting users' net switching costs. 
As shown in \textrm{F}ig. 3, the large operator 1 charges a high price while the small operator 2 charges a low price.
Thus, the asymmetries of their market shares and prices fade away over time, resulting in symmetric outcomes in steady state. 
We also observe that the slower rate of the steady state as $T$ increases. 
 
 \begin{figure}[t]
\centering
\includegraphics[angle=0,width=3.1in]{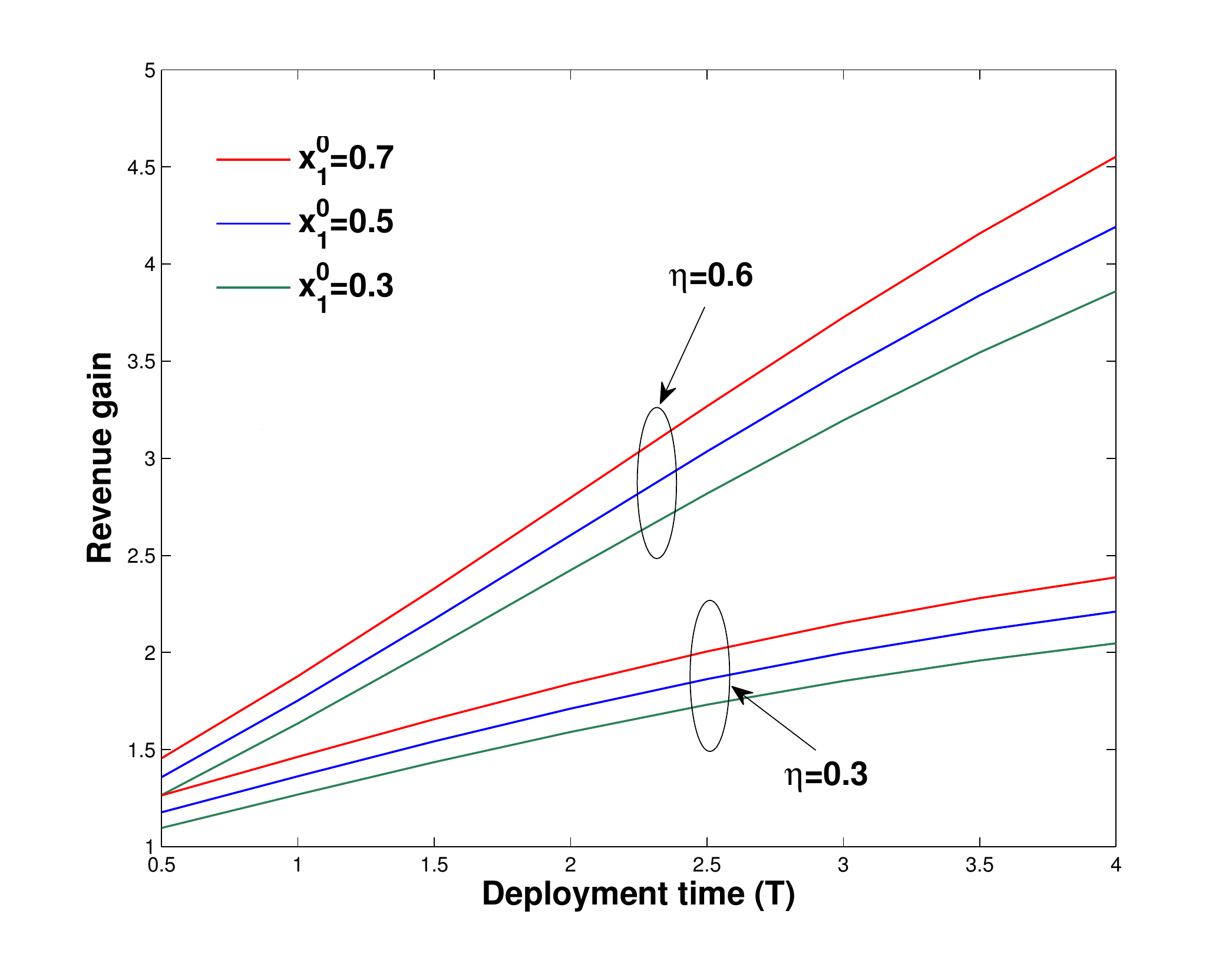}
\caption{Revenue gain as a function of $T$ under different parameters $\bold{x}(0)$ and $\eta$.
Other parameters are $u_{0}=10$, $\rho=0.5$, $\underline{s}=5$, and $\overline{s}=10$.}
\vskip 5pt
\end{figure}

\subsection{Values of Contiguous and Non-contiguous Spectrum}
In this subsection, 
we examine the values of two different spectrum blocks: contiguous spectrum block $A$ and non-contiguous spectrum block $B$. 
To this end, we define the revenue gain as follows: 
\begin{eqnarray}
\mathcal{R}_{gain}=\frac{\mathcal{R}_{i}^{A}}{\mathcal{R}_{j}^{B}}, \quad i=1,2, \quad i\ne j.
\end{eqnarray}
\textrm{F}ig. 4 shows how $\mathcal{R}_{gain}$ changes in $T$ with different parameters $\eta$ and $\bold{x}(0)$.
Intuitively, the revenue gain is strictly increasing over $T$ and this gain becomes much higher as $\eta$ or $x_{1}^{0}$ increase. 
It explains how critical new spectrum is allocated to both operators, and why they should spitefully bid in a first-price sealed-bid auction to 
achieve a dominant position or compensate the revenue gap. 
It also indicates that 
the asymmetric-valued LTE spectrum allocation should be carefully tailored to 
avoid excessive competition, and promote fair competition between operators.

\subsection{Bidding Strategies and Resultant Profits}
Based on the estimated values of two different spectrum blocks, 
we investigate the two operators' equilibrium bidding strategies and their corresponding profits. 
To this end, the initial market share state are set to $\bold{x}(0)=[0.6, 0.4]$.
\textrm{F}ig. 5 shows how the two operator's equilibrium bidding strategies realize when the spite coefficient $\gamma$ varies under two different 
deployment times.
Intuitively, the more spiteful the two operators are, the more aggressively they tend to bid. 
Since more spiteful operator gets more disutility from the profit of his competitor, 
he is willing to sacrifice some monetary payoff in order to minimize his rival's profit.
It is worth noting that there is a cross-over point at $\gamma=0.5$. 
This can be interpreted as the different levels of the utility (or disutility) of winning (or losing) the auction between the two operators.
When $\gamma < 0.5$, the more emphasis they put on their own profits, inducing operator 1 to place a higher bid
because he could make more profit from $A$. 
When $\gamma > 0.5$, on the other hand, they are more interested in minimizing the profits of their competitors, 
prompting operator 2 to bid more aggressively since she could reduce operator 1's profit more when acquiring $A$, and thus increase her relative position in the market. This may explains why Korea Telecom (i.e., a small operator) placed a bid twice as high as SK Telecom (i.e., a large operator) for procuring the contiguous spectrum block \cite{Moody:2013}.
We also observe that the equilibrium bidding strategies of the two operator increase as $T$ increases. 
It implies that the longer $T$ promotes the more fierce competition between operators.

\begin{figure}[t]
\centering
\includegraphics[angle=0,width=3.1in]{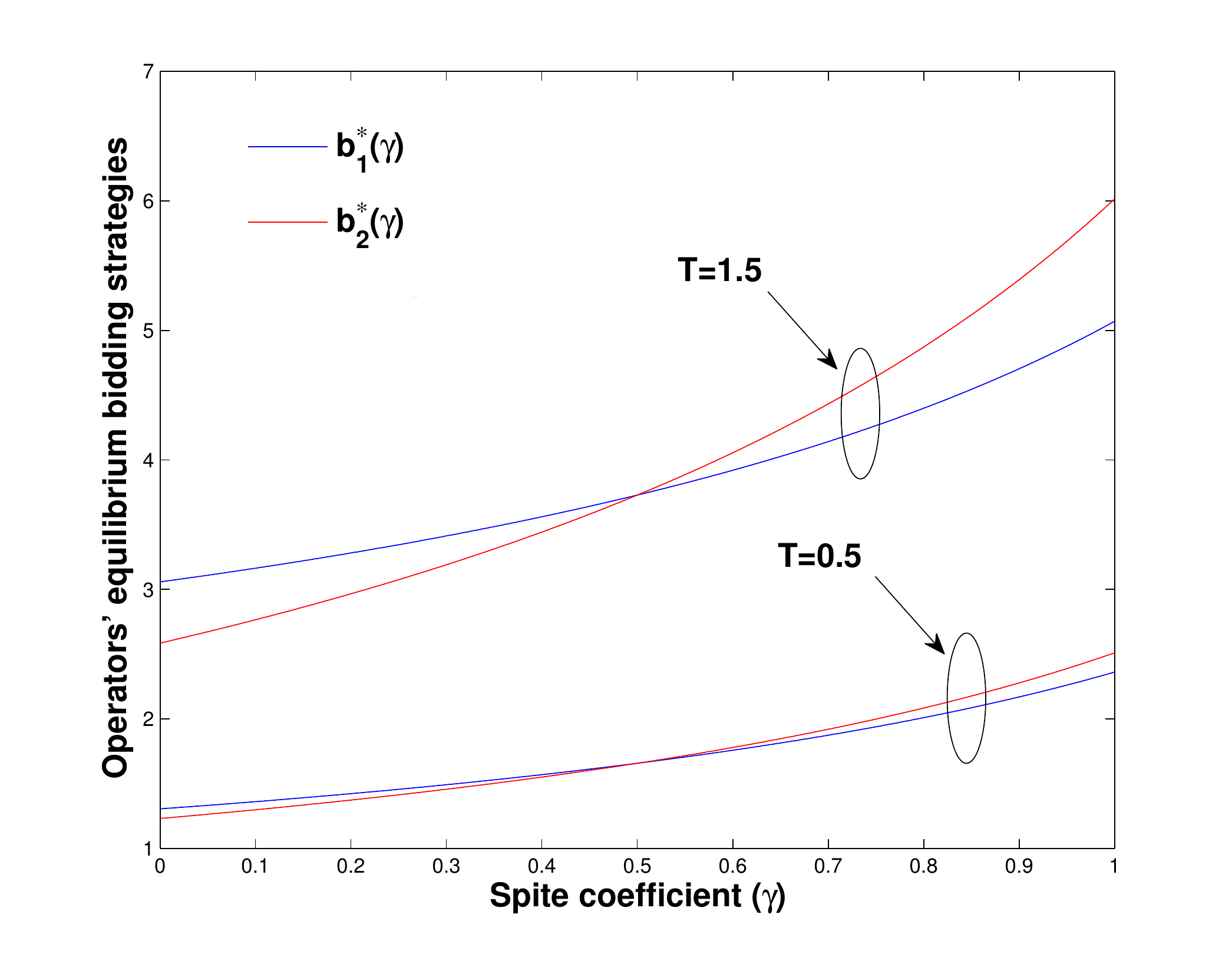}
\caption{The equilibrium bidding strategies of two operators as a function of $\gamma$ under two different deployment times.
Other parameters are $u_{0}=10$, $\eta=0.5$, $\rho=0.5$, $\underline{s}=5$, $\overline{s}=10$, $c^{A}=0.1$, $c^{B}=0.2$, and $c^{BS}=1$.}
\vskip 5pt
\end{figure}

\textrm{F}ig. 6 shows the resultant profits of the two operators as a function of $\gamma$ under two different deployment times.
When $\gamma<0.5$, the two operators are more interested in maximizing their profits, 
resulting in the large difference between their profits as $T$ increases or $\gamma$ decreases.
It implies that the small operator 2 should place a more spiteful bid as $T$ increases in order to improve (or maintain) his own standing in a highly  competitive market. 
When $\gamma>0.5$, the difference between their profits decreases as $T$ decreases for the same $\gamma$, allowing operator 2 to 
increase his relative position in the market by bidding more aggressively. 
However, when $\gamma=0.6$ and $T=1.5$, the profit of operator 2 is is more than 35\% compared with that of operator 1.
It offers meaningful insights into the role of the regulators in that the longer $T$ hinders fair competition between the two operators. 
Thus, the regulators should appropriately impose limits on the timing of the double-speed LTE services.
In South Korea, for example, Korea Telecom (KT) who secured the contiguous spectrum 
could not start its double-speed LTE services in major cities until March 2014, and nation-wide coverage until July \cite{Moody:2013}.
This gave its competitors time to roll out the LTE-Advanced services, allowing to promote fair competition between operators. 
\vskip -5pt
\begin{figure}[ht]
\centering 
 {
   \includegraphics[angle=0,width=2.8in] {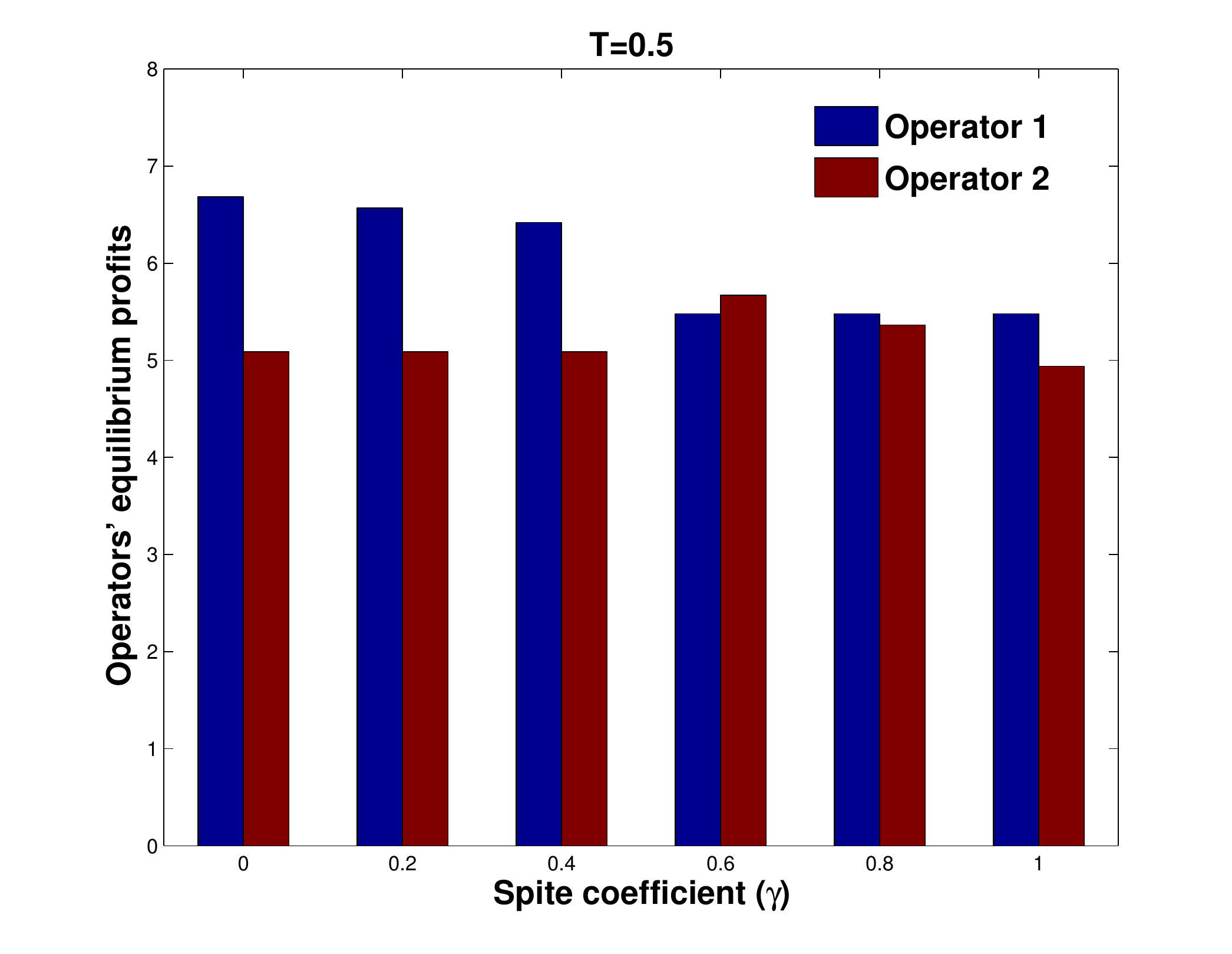}
 }
\vskip -5pt
{
   \includegraphics[angle=0,width=2.8in] {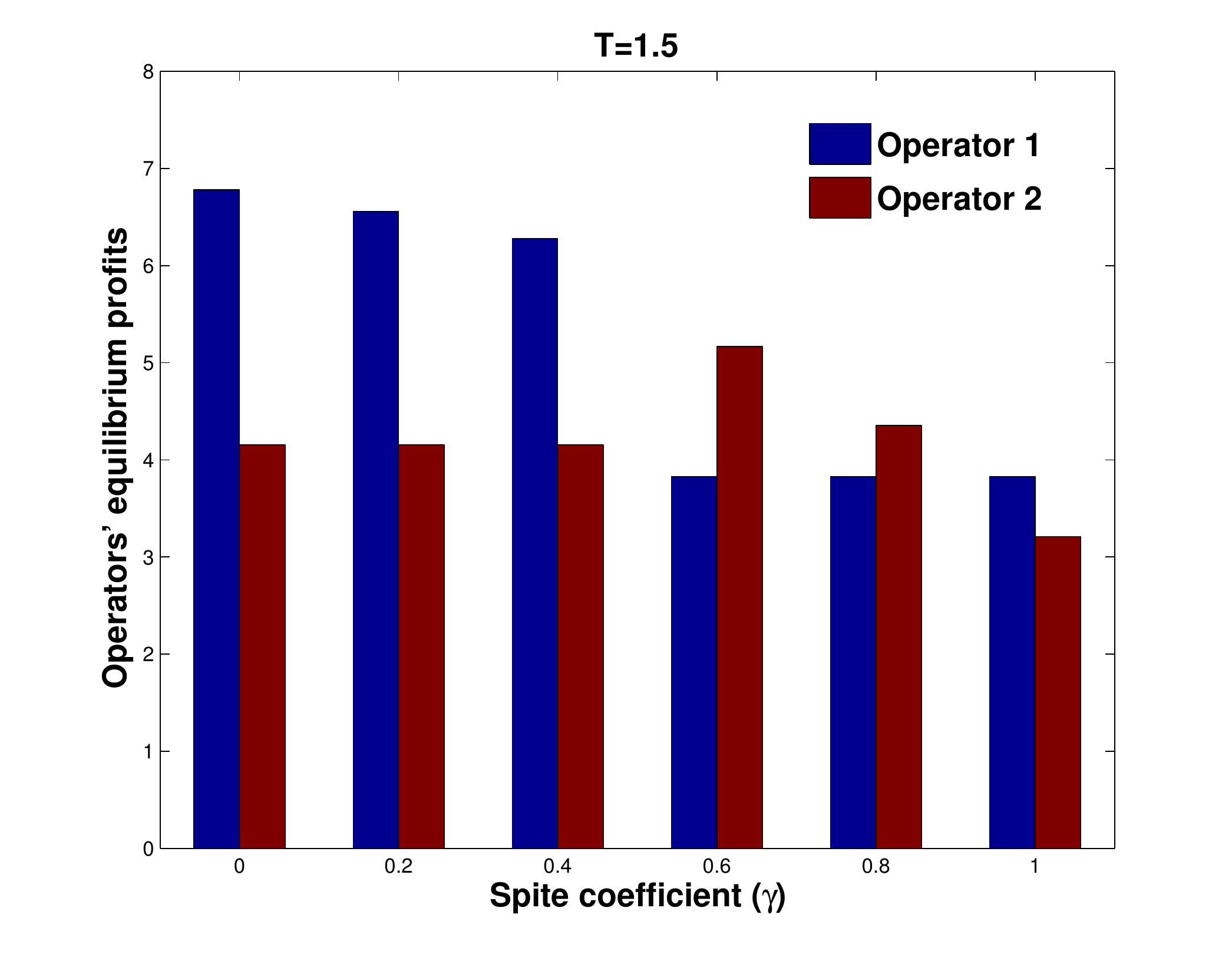}
 }
\caption{The equilibrium profits of two operators as a function of $\gamma$ under two different deployment times. $u_{0}=10$, $\eta=0.5$, $\rho=0.5$, $\underline{s}=5$, $\overline{s}=10$, $c^{A}=0.1$, $c^{B}=0.2$, and $c^{BS}=1$.}
\vskip 5pt
\end{figure}

\section{Conclusion}
This paper presents a comprehensive analytical and numerical studies of the effects of asymmetric-valued LTE spectrum allocation.
We model the interactions between operators and users as a hierarchical dynamic game framework, where two spiteful operators
simultaneously make spectrum acquisition decisions in the upper-level first-price seals-bid auction game, 
and dynamic pricing decisions in the lower-level differential game, taking into account user subscription dynamics. 
Using backward induction, we derive and characterize the equilibrium of the entire game under mild conditions. 
Through analytical and numerical results, we verify our studies by comparing the latest results of LTE spectrum auction in South Korea. 
This serves as the benchmark of asymmetric-valued LTE spectrum allocation.

The weakness of this study is the same spite between two operators. 
Depending on the relative position in the market, operators may take the profits of them into account differently. 
This broader setting where operators are asymmetric spiteful makes our studies more applicable for extensive spectrum allocation scenarios 
and gives more insights into the role of regulation.

\section{Appendix}
\subsection{Proof of Proposition 1}
From (12), we have the two operators' best response functions, 
\begin{equation}\small
p_{1}=\frac{1}{2}\left(p_{2} + \eta u_{0} + \underline{s} - \lambda_{1} + s_{}x_{1}\right), p_{2}=\frac{1}{2}\left(p_{1}  - \eta u_{0} + \overline{s} + \lambda_{2} - s_{2}x_{1}\right].
\end{equation} \normalsize
Using them gives equilibrium prices, 
\begin{equation}\small
p_{1}=\frac{1}{3}\left( \eta u_{0} + s_{1}+\underline{s} - 2\lambda_{1} + \lambda_{2} + s_{2}x_{1} \right),\nonumber
\end{equation} \normalsize
\begin{equation}\small
p_{2}=\frac{1}{3}\left( -\eta u_{0} +s_{1} + \overline{s} - \lambda_{1} + 2\lambda_{2} - s_{2}x_{1}\right).
\end{equation} \normalsize
Next, from (13) and (35), we obtain the following partial differential equations, 
\begin{equation}\small
\dot{\lambda}_{1}=\rho\lambda_{1}-\frac{2p_{1}s_{2} - \lambda_{1}(5\underline{s}+\overline{s})}{3s_{1}} ,
\dot{\lambda}_{2}=\rho \lambda_{2}-\frac{-2p_{2}s_{2} - \lambda_{2}(5\underline{s}+\overline{s})}{3s_{1}} . 
\end{equation} \normalsize
We solve this system of differential equations by the method of undetermined coefficients.
Assume that $\lambda_{i}(t)=k_{i}(t)x_{1}(t) + e_{i}(t)$ for $i=1, 2$.
Substitution into (36) yields 
\begin{equation}\small
\dot{k}_{1}x_{1}+k_{1}\dot{x}_{1} +\dot{e}_{1}=\rho \left(k_{1}x_{1}+e_{1}\right)- \frac{2p_{1}s_{2} - (k_{1}x_{1}+e_{1})(5\underline{s}+\overline{s})}{3s_{1}}, 
\end{equation} \normalsize
\begin{equation}\small
\dot{k}_{2}x_{1}+k_{2}\dot{x}_{1} + \dot{e}_{2}=\rho \left(k_{2}x_{1}+e_{2}\right)-\frac{-2p_{2}s_{2}-(k_{2}x_{1}+e_{2})(5\underline{s}+\overline{s})}{3s_{1}},
\end{equation} \normalsize
where $\dot{x}_{1}$ is given in (8). 
Since the equalities in  (37) and (38) must hold for all values of $x_{1}$, the coefficients of $x_{1}$ and the constant terms have to be zero. 
Define $y(t)=k_{1}(t)-k_{2}(t)$ and $z(t)=e_{1}(t)-e_{2}(t)$. 
Subtracting (38) from (37) yields 
\begin{equation}\small
\dot{y} - \left(\frac{k_{1}+k_{2}+7\underline{s}+5\overline{s}+3\rho s_{1}}{3s_{1}}\right)y=0, \quad \dot{z} - (1+\rho)z + \frac{2s_{2}}{3}=0.
\end{equation} \normalsize
Using the integrating factor method under the boundary conditions $\lambda_{i}(T)=0$ for $i=1,2$, which implies $k_{i}(T)=e_{i}(T)=0$ for $i=1,2$, 
we obtain 
\begin{equation}\small
y(t)=0, \quad z(t)=\frac{2s_{2}}{3(1+\rho)}\left(1- e^{(1+\rho)(t-T)}\right), 
\end{equation} \normalsize
which follows that $k_{1}(t)=k_{2}(t)$ and $e_{1}(t)=e_{2}(t)+z(t)$.

With the above results, setting $k(t)=k_{1}(t)=k_{2}(t)$ and rewriting (37) and (38) give
\begin{equation}\small
\dot{k}=-\frac{2}{3s_{1}}k^{2}+ \left(\frac{11\overline{s}+ 25\underline{s}+9\rho s_{1})}{9s_{1}}\right)k - \frac{2}{9}s_{2}, \\
\end{equation} \normalsize
\begin{equation}\small
\dot{e}_{1}=\left(\frac{13\underline{s}+5\overline{s}+6k}{9s_{1}}+\rho\right)e_{1} - \frac{(2s_{2}+3k)(s_{1}+\underline{s}+\eta u_{0}-z)}{9s_{1}}.
\end{equation} \normalsize
First, consider the Riccati differential equation (41). 
Let $\alpha_{1}$ and $\alpha_{2}$ are the two solutions of the quadratic equation, i.e., 
\begin{equation}\small
\frac{2}{3s_{1}}k^{2} - \left(\frac{11\overline{s}+25\underline{s}+9\rho s_{1}}{9s_{1}}\right)k + \frac{2}{9}s_{2}=0.
\end{equation} \normalsize
Without loss of generality, assume $\alpha_{1}<\alpha_{2}$. Note that the two solutions $\alpha_{1}$ and $\alpha_{2}$ 
are particular solutions of (41). 
Then a general solution of (41) is (see \cite{Dockner:2000})
\begin{equation}\small
\frac{k(t)-\alpha_{1}}{k(t)-\alpha_{2}}=Ce^{-\frac{2(\alpha_{1}-\alpha_{2})}{3s_{1}}t},
\end{equation} \normalsize
where $C$ is the constant of integration.
Using the boundary condition $k(T)=0$, the constant $C$ is $\frac{\alpha_{1}}{\alpha_{2}}e^{\frac{2(\alpha_{1}-\alpha_{2})}{3s_{1}}T}$, which yields
\begin{equation} \small
k(t)=\frac{\alpha_{1}\left(1- e^{\frac{2(\alpha_{1}-\alpha_{2})}{3s_{1}}(T-t)}\right)}{1- \frac{\alpha_{1}}{\alpha_{2}}e^{\frac{2(\alpha_{1}-\alpha_{2})}{3s_{1}}(T-t)}}.
\end{equation} \normalsize

Next, consider the first order partial differential equation (42). 
Let $\mu(t)=\frac{13\underline{s}+5\overline{s}+6k(t)}{9s_{1}}+\rho$ and $\nu(t)=\frac{\left(2s_{2}+3k(t)\right)\left(z(t)- 2\underline{s}- \overline{s} -\eta u_{0} \right)}{9s_{1}}$.
Using the integrating factor method under the boundary conditions $e_{i}(T)=0$ for $i=1, 2$, we have
\begin{equation} \small
e_{1}(t)=-e^{\int_{0}^{t}\mu(\xi)d\xi}\left(\int_{t}^{T}e^{-\int_{0}^{\tau}\mu(\xi)d\xi}\nu(\tau)d\tau\right), e_{2}(t)=e_{1}(t)-z(t), 
\end{equation} \normalsize

Then the equilibrium prices can be rewritten as
\begin{equation}\small
p_{1}^{\ast}(t,x_{1}(t))=\frac{\eta u_{0} + \overline{s} +2\underline{s} - e_{1}(t) - z(t)}{3} + \frac{s_{2}-k(t)}{3}x_{1}(t),
\end{equation}\normalsize
\begin{equation}\small
p_{2}^{\ast}(t, x_{1}(t))=\frac{-\eta u_{0} + 2\overline{s} +\underline{s} + e_{2}(t)- z(t)}{3} - \frac{s_{2}-k(t)}{3}x_{1}(t).
\end{equation}\normalsize
Now, operator 1's market share rate in (8) can be expressed as
\begin{equation}\small
\dot{x}_{1}=\frac{\eta u_{0} +2\underline{s} + \overline{s} +e_{1}+ e_{2}}{3s_{1}}+ \frac{2(k-\overline{s}-2\underline{s})}{3s_{1}}x_{1},
\end{equation} \normalsize
where his initial market share is $x_{1}^{0}$.
Using the integrating factor method yield
\begin{equation} \small
x_{1}^{\ast}(t)=e^{\int_{0}^{t}\delta(\xi)d\xi}\left(x_{1}^{0}+ \int_{0}^{t}e^{-\int_{0}^{\tau}\delta(\xi)d\xi}\zeta(\tau)d\tau\right),
\end{equation} \normalsize
where $\delta(t)=\frac{2(k(t)-b-2a)}{3(a+b)}$ and $\zeta(t)=\frac{\eta u_{0} +2a + b +e_{1}(t)+ e_{2}(t)}{3(a+b)}$,
which completes the proof.

\subsection{Proof of Proposition 2}
We proceed by following the similar steps as in Appendix A. 
The Hamiltonian function for operator $i$ is
\begin{equation}
\mathcal{H}_{i}=\left(r_{i}(t) + \lambda_{i}(t)\dot{x}_{1}(t)\right), \quad\ i=\{1, 2\},
\end{equation}
Due to the concavity of Hamiltonian function with respect to $p_{i}$, the necessary conditions 
for the feedback Nash equilibrium provide sufficient conditions for optimality, i.e., 
\setlength\arraycolsep{2pt}
\begin{eqnarray}
\dot{x}_{1}&=&\frac{p_{2}-p_{1}+\underline{s}}{s_{1}} - \frac{2\underline{s}}{s_{1}}x_{1}, \quad x_{1}(T)=x_{1}^{T},\\
\frac{\partial \mathcal{H}_{i}}{\partial p_{i}}&=&0, \\
\frac{\partial \lambda_{i}}{\partial t}&=&\rho \lambda_{i}-\frac{\partial \mathcal{H}_{i}}{\partial x_{1}} - \frac{\partial \mathcal{H}_{i}}{\partial p_{j}}\frac{\partial p_{j}}{\partial x_{1}},\\
\lim_{t \to \infty}e^{-\rho t}\lambda_{i}(t)&=&0.
\end{eqnarray}
Note that the terminal condition in (55) is different from (14).

From (53), the equilibrium prices are
\begin{equation}\small
p_{1}=\frac{1}{3}\left(s_{1}+\underline{s} - 2\lambda_{1} + \lambda_{2} + s_{2}x_{1}\right), 
p_{2}=\frac{1}{3}\left(s_{1}+\overline{s} - \lambda_{1} + 2\lambda_{2} - s_{2}x_{1}\right).
\end{equation} \normalsize
From (54) and (56), we have
\begin{equation} \small
\dot{\lambda}_{1}=\rho \lambda_{1}-\frac{2p_{1}s_{2} - \lambda_{1}(5\underline{s}+\overline{s})}{3s_{1}},
\dot{\lambda}_{2}=\rho \lambda_{2}-\frac{-2p_{2}s_{2} - \lambda_{2}(5\underline{s}+\overline{s})}{3s_{1}}. 
\end{equation} \normalsize
 
We solve this system of differential equations by the method of undetermined coefficients.
Assume that $\lambda_{i}(t)=k_{i}x_{1}(t) + e_{i}$ for $i=1, 2$. Note that the coefficients of $x_{1}(t)$ and the constant terms
are not functions of time when the time horizon is infinite. 
Substitution into (47) yields 
\begin{equation} \small
k_{1}\dot{x}_{1} =\rho(k_{1}x_{1}+e_{1})-\frac{2p_{1}s_{2} -(k_{1}x_{1}+e_{1})(5\underline{s}+\overline{s})}{3s_{1}},
\end{equation} \normalsize
\begin{equation} \small
k_{2}\dot{x}_{1}=  \rho(k_{2}x_{1}+e_{2})+\frac{2p_{1}s_{2} +(k_{2}x_{1}+e_{2})(5\underline{s}+\overline{s})}{3s_{1}}.
\end{equation} \normalsize
Subtracting (59) from (58), as in the proof of Proposition 1, we can show that 
\begin{equation}\small
k_{1}=k_{2}=\frac{11\overline{s}+25\underline{s}+9\rho s_{1}-\sqrt{\left(11\overline{s}+25\underline{s}+9\rho s_{1}\right)^{2} -48(s_{2})^{2}}}{12}, \nonumber
\end{equation} \normalsize
\begin{equation}\small
e_{1}=e_{2}+ \frac{2s_{2}}{3(1+\rho)}.
\end{equation} \normalsize
where $k_{1}$ (or $k_{2}$) is the smallest root of the quadratic equation 
\begin{equation}\small
6k^{2}-\left(11\overline{s}+25\underline{s}+9\rho s_{1}\right)k + 2s_{2}^{2}=0.\nonumber
\end{equation} \normalsize

Let $k=k_{1}=k_{2}$. Using (58), (59), and (60) we have
\begin{equation}\small
e_{1}=\frac{s_{2}}{3(1+\rho)}+ \frac{s_{2}^{2}-3k(2\underline{s}+\overline{s})}{6k -13\underline{s}-5\overline{s}-9\rho s_{1}}, k=-e_{1}-e_{2}.
\end{equation} \normalsize
Then, the equilibrium prices can be rewritten as 
\begin{equation}\small
p_{1}^{\ast}(t,x_{1}(t))=\frac{s_{1} + \underline{s}+e_{2}-2e_{1}}{3} + \frac{s_{2}-k}{3}x_{1}(t),
\end{equation} \normalsize
\begin{equation}\small
p_{2}^{\ast}(t,x_{1}(t))=\frac{s_{1}+\overline{s}+2e_{2}-e_{1}}{3} - \frac{s_{2}-k}{3}x_{1}(t).
\end{equation} \normalsize

Following similar steps in (49) and (50), user subscription dynamics in symmetric phase can be written as 
\begin{equation} \small
x_{1}^{\ast}(t)=\frac{1}{2} + \left(x_{1}^{T} - \frac{1}{2}\right)e^{\frac{2(k-\underline{s}-s_{1})}{3s_{1}}(t-T)}, \quad x_{2}^{\ast}(t)=1-x_{1}^{\ast}(t),
\end{equation} \normalsize
which completes the proof.

\subsection{Proof of Proposition 3}
Without loss of generality, suppose that operator $i$ knows his bid $b_{i}$. 
Further, we assume that operator $i$ infer that the bidding strategy of operator $j$ on $A$ is 
drawn uniformly and independently from $[c^{A}, \mathcal{R}_{j}^{A}]$. The operator $i$'s optimization problem is to choose $b_{i}$ to maximize
the expectation of 
\setlength\arraycolsep{-1pt}
\begin{eqnarray}
\mathrm{E}_{b_{i}}(\Pi_{i})=&\;&\int_{c^{A}}^{b_{i}}\left[(1-\gamma)(\mathcal{R}_{i}^{A} - b_{i}) -\gamma \pi_{j}^{B}\right]f(b_{j})db_{j}\nonumber\\
&+&\int_{b_{i}}^{\mathcal{R}_{j}^{A}}\left[(1-\gamma)\pi_{i}^{B} -\gamma\left(\mathcal{R}_{j}^{A}- b_{j}\right)\right]f(b_{j})db_{j}.
\end{eqnarray}
Differentiating (65) with respect to $b_{i}$, 
setting the result to zero and multiplying by $\mathcal{R}_{j}^{A} - c^{A}$ give
\setlength\arraycolsep{2pt} \thickmuskip=-1mu\medmuskip=-1mu \thinmuskip=1mu
\begin{eqnarray}
\frac{\partial \mathrm{E}_{b_{i}}(\Pi_{i})}{\partial b_{i}}=(1-\gamma)\left(\mathcal{R}_{i}^{A}+c^{A}-\pi_{i}^{B}\right)+\gamma\left(\mathcal{R}_{j}^{A}-\pi_{j}^{B}\right)-(2-\gamma)b_{i}=0.\nonumber\\
\end{eqnarray}
Applying the same way to operator $j$ completes the proof.

\end{document}

%% file: header.tex
\newtheorem{theorem}{Theorem}

\newtheorem{definition}[theorem]{Definition}

\def\({\left(}
\def\){\right)}

\def\[{\left[}
\def\]{\right]}



%% file: JungKim_JCN.bbl
\begin{thebibliography}{99}

\bibitem{Cisco:2014}
Cisco,
\newblock ``Cisco visual networking index: global mobile data traffic
forecast update, 2014-2019,"
\newblock {\it White paper},
\newblock Feb. 
\newblock 2015.

\bibitem{America:2013}
4G Americas,
\newblock ``Meeting the 1000x challenge: The need for spectrum, technology and policy innovations,''
\newblock {\it White paper}, 
\newblock Oct. 
\newblock 2013.

\bibitem{Huang:2013}
J. Huang and L. Gao, 
\newblock {\it Wireless Network Pricing}, 
\newblock Synthesis Lectures on Communication Networks,
\newblock Morgan \& Claypool,
\newblock 2013. 

\bibitem{Marcus:2015}
M. J. Marcus, 
\newblock ``Spectrum policy and wireless innovation,''
\newblock {\it IEEE Wireless Commun.},
\newblock vol. 22, 
\newblock no. 1,
\newblock pp. 8--9,
\newblock Feb. 
\newblock 2015. 

\bibitem{Mikio:2010}
M. Iwamura {\it et al.},  
\newblock ``Carrier aggregation framework in 3GPP LTE-advanced,''
\newblock {\it IEEE Commun. Mag.},
\newblock vol. 48,
\newblock no. 8, 
\newblock pp. 60--67,
\newblock Aug.
\newblock 2010,



\bibitem{Moody:2013}
Moody's Investor Service,
\newblock ``Sector comment: Spectrum auction results are credit positive for major Korean telcos,''
\newblock Sep.
\newblock 2013. 




\bibitem{Milgrom:1982}
P. R. Milgrom and R. J. Weber,
\newblock ``A theory of auctions and competitive bidding,"
\newblock {\it Econometrica},
\newblock vol. 50,
\newblock no.5,
\newblock pp. 1089--1122,
\newblock Sep.
\newblock 1982.

\bibitem{Jia:2008}
J. Jia and Q. Zhang,
\newblock ``Competitions and dynamics of duopoly wireless service providers in dynamic spectrum market,"
\newblock in {\em Proc. ACM MobiHoc},
\newblock (Hong Kong SAR, China),
\newblock May 2008,
\newblock pp. 313--322. 

\bibitem{Duan:2011}
L. Duan, J. Huang, and B. Shou,
\newblock ``Duopoly competition in dynamic spectrum leasing and pricing,"
\newblock {\it IEEE Trans. Mobile Compu.},
\newblock vol. 11,
\newblock no.11,
\newblock pp. 1706--1719,
\newblock Nov.
\newblock 2012.



\bibitem{Feng:2012}
X. Feng, Y. Chen, J. Zhang, Q. Zhang, and B. Li,
\newblock ``TAHES: A truthful double auction mechanism for heterogeneous spectrums,''
\newblock {\it IEEE Trans. Wireless Commun.},
\newblock vol. 11, 
\newblock no. 11, 
\newblock pp. 4038--4047,
\newblock Nov.
\newblock 2012.

\bibitem{Jung:2013}
S. Y. Jung, S. M. Yu, and S.-L. Kim,
\newblock ``Utility-optimal partial spectrum leasing for future
wireless services,"
\newblock in {\em Proc. IEEE VTC},
\newblock (Dresden, Germany),
\newblock Jun.
\newblock 2013.

\bibitem{Yu:2014}
S. M. Yu and S.-L. Kim, 
\newblock ``Game-theoretic understanding of price dynamics in mobile communication services,''
\newblock {\it IEEE Trans. Wireless Commun.},
\newblock vol. 13, 
\newblock no. 9, 
\newblock pp. 5120--5131, 
\newblock Sep.
\newblock 2014. 


\bibitem{Illing:2003}
G. Illing and U. Kluh, 
\newblock {\it Spectrum Auctions and Competition in Telecommunications},
\newblock MIT Press,
\newblock 2004.


\bibitem{Morgan:2003}
J. Morgan, K. Steiglithz, and G. Reis,
\newblock ``The spite motive and equilibrium behavior in auctions,"
\newblock {\it Contrib. Econ. Anal. Pol.},
\newblock vol. 2,
\newblock no. 1
\newblock pp. 1102--1127,
\newblock 2003.


\bibitem{Jung:2014}
S.Y. Jung, S. M. Yu, and S. L. Kim, 
\newblock ``Asymmetric-valued spectrum auction and competition in wireless broadband services,''
\newblock in {\em Proc. IEEE WiOpt},
\newblock 2014.

\bibitem{Chau:2010}
C.-K. Chau, Q. Wang, and D.-M. Chiu,
\newblock ``On the viability of paris metro pricing for communication and service networks,''
\newblock in {\em Proc. IEEE INFOCOM},
\newblock (San Diego, USA),
\newblock Mar. 2010, 
\newblock pp. 929--937.


\bibitem{Ren:2011}
S. Ren, J. Park, and M. van der Schaar, 
\newblock ``User subscription dynamics and revenue maximization in communications markets,''
\newblock in {\em Proc. IEEE INFOCOM}, 
\newblock (Shanghai, China),
\newblock Apr. 2011,
\newblock pp. 2696--2704.


\bibitem{ITU:statistics}
ITU, 
\newblock ``The world in 2014: ICT facts and figures,'' 
\newblock Apr.
\newblock 2014.


\bibitem{Shin:2014}
M. Shin, 
\newblock ``How the intense competition among its telco players makes Korea the leading nation for mobile,''
\newblock Available: http://www.atelier.net,
\newblock Apr.
\newblock 2014.



\bibitem{Han:2011}
Z. Han \textit{et al.}, 
\newblock {\it Game Theory in Wireless and Communication Networks: Theory, Models, and Applications},
\newblock Cambridge Univ. Press,
\newblock 2011.


\bibitem{Dockner:2000}
E. J. Dockner, S. Jorgensen, N. V. Long, and G. Sorger, 
\newblock {\it Differential Games in Economics and Management Science},
\newblock Cambridge Univ. Press, 
\newblock 2000.

\end{thebibliography}
